# Exploring the Factors of "AI Guilt" Among Students – Are You Guilty of Using AI in Your Homework?

Cecilia Ka Yuk Chan
The University of Hong Kong
Email: ckchan09@hku.hk

**Abstract**
This study explores the phenomenon of "AI guilt" among secondary school students, a form of moral discomfort arising from the use of AI tools in academic tasks traditionally performed by humans. Through qualitative methodologies, the research examines the factors contributing to AI guilt, its social and psychological impacts, and its implications for educational practices. The findings revealed three main dimensions for AI guilt - perceived laziness and authenticity, fear of judgment, and identity and self-efficacy concerns. The findings suggest a need to redefine academic integrity and shift our mindset to reconsider what we should value in education. The study also emphasizes the importance of ethical guidelines and educational support and provides implications to help students navigate the complexities of AI in education, reducing feelings of guilt while enhancing learning outcomes.

**Introduction**
In the advancing field of artificial intelligence (AI), algorithms can now assist in writing essays (Lim et al., 2023; Swiecki et al., 2022; Chan & Hu, 2023), generating presentation slides (Baidoo-Anu & Ansah, 2023; Feuerriegel et al., 2024), designing and creating personalised assessments (Chiu, 2023; Hodges & Kirshner, 2023; Moorhouse et al., 2023) and even conducting research (Chan & Colloton, 2024, p) in education. This integration of AI in academic settings has introduced an emotional response among students known as "AI guilt". This phenomenon, deeply intertwined with our intrinsic values and ethics, reflects the complex relationship between human intellect and machine capabilities; creating an internal conflict and ethical dilemmas students experience when using AI for tasks traditionally requiring human intellect and creativity. AI Guilt prompts students and educators to question the essence of learning, originality, and personal effort.

Empirical studies reveal mixed feelings among students regarding the use of AI tools. Paulise (2024) reported that 36% of Generation Z individuals felt guilty about using AI tools like ChatGPT for work tasks, with concerns about undermining personal effort and the authenticity of their outputs. Similarly, Chan and Hu (2023) found that students recognize the benefits of generative AI but express concerns about ethical implications and the potential for overreliance.

Attewell (2023) conducted a study involving over 200 students and discovered that students increasingly view generative AI as a collaborative tool rather than a mere answer provider. However, they also highlighted concerns about equity, bias, and the ethical implications of AI use. Chan and Lee (2023) noted generational differences in AI adoption, with Gen Z students showing more interest in using AI for educational purposes compared to Gen X and millennial teachers.

The objective of this study is to explore the current perceptions of AI guilt among secondary school students, investigating the reasons behind these feelings and their implications for educational practices. By examining how students navigate the ethical landscape of AI usage, the study aims to provide insights that can inform the creation of guidelines and policies to support ethical AI integration in education. This research will address critical questions about the balance between leveraging AI's capabilities and maintaining the intrinsic value of human creativity and effort, ultimately contributing to a more ethically grounded and effective educational system.

**Defining AI Guilt**
AI Guilt is a relatively new concept that refers to the moral or ethical discomfort individuals feel when using AI tools for tasks traditionally performed by humans. This feeling of guilt may come from concerns about originality, authenticity, and the perception of laziness or cheating. We therefore defined

> ***AI Guilt*** *as a psychological phenomenon wherein individuals feel guilt or moral discomfort when use AI tools and feel that their reliance on these technologies might be perceived negatively by others or feel disingenuous. This could be particularly relevant in academic, professional, or creative contexts where originality and individual effort are highly valued.*

Research on the psychological impacts of AI in educational contexts is still emerging. Studies indicate that while AI can enhance learning and productivity, it can also induce stress and anxiety, particularly when students feel that their skills are being undermined (Chan & Tsi, 2024; Kim et al, 2023). The ethical implications of AI use in education are equally significant. Holmes & Miao (2023) state that ethical guidelines are essential to ensure that AI contributes positively to educational outcomes without compromising ethical standards.

**Rationale for Understanding Students' AI Guilt**
Evaluating students' AI guilt is a critical endeavor in the modern educational landscape, where AI is increasingly integrated into academic work. One of the key reasons for evaluating AI guilt is to gauge the extent to which students feel that using AI tools diminishes the authenticity of their work. Many students may experience a conflict between the ease and efficiency provided by AI and their own intrinsic values related to originality and personal effort. This conflict can lead to feelings of guilt, where students question whether their achievements are genuinely their own.

The Technology Acceptance Model (TAM) by Davis (1989) provides foundational insights into how individuals accept and use technology, focusing on perceived usefulness and ease of use. When students feel that AI tools are useful and easy to use, yet conflict with their intrinsic values of originality and effort, it can result in significant cognitive dissonance. Studies on cognitive dissonance, such as Festinger's (1957) theory, examine the mental discomfort experienced when holding conflicting beliefs or attitudes, may play a critical role here.

Moreover, evaluating AI guilt can shed light on the social dynamics within educational settings. Students may worry about being judged by their peers or instructors for using AI tools, perceiving such use as a form of cheating or taking shortcuts. The Unified Theory of Acceptance and Use of Technology (UTAUT) developed by Venkatesh et al. (2003)

highlights the influence of social factors, including peer and social pressure, on technology adoption. This social pressure can exacerbate feelings of guilt and discourage the use of potentially beneficial technologies. By assessing these concerns, educators can foster a more supportive environment that encourages ethical use of AI and reduces stigma.

Additionally, understanding AI guilt is crucial for developing comprehensive ethical guidelines and educational policies. The Technological Pedagogical Content Knowledge (TPACK) framework, proposed by Mishra and Koehler (2006), emphasizes the integration of technology in ways that align with pedagogical goals and content knowledge. Clear teacher guidelines based on this framework can help clarify what constitutes acceptable use of AI in academic work, thereby reducing uncertainty and guilt among students. For instance, if students know that using AI for certain tasks is explicitly allowed and even encouraged, they may feel more comfortable and confident in integrating these tools into their work. Academic integrity concepts are essential here, as the use of AI tools raises questions about the authenticity and originality of students' work.

Evaluating AI guilt also provides insights into how students perceive the balance between technological advancement and maintaining personal skill sets. As AI tools become more advanced, there is a risk that students might rely too heavily on them, potentially at the expense of developing critical thinking and problem-solving skills. By understanding the extent of AI guilt, educators can better design curricula that integrate AI in a way that enhances learning while also promoting the development of essential skills. Impostor Syndrome (Clance and Imes, 1978), characterized by feelings of inadequacy despite evident success, can intersect with AI guilt as students may feel their achievements are not truly their own when assisted by AI. Addressing these feelings is essential to help students use AI responsibly and ethically, without feeling that they are undermining their own capabilities or the value of their work.

**Research Questions**

1. What are the factors that lead to AI guilt among secondary school students?
2. What are the social and psychological impacts of AI guilt on secondary school students?
3. How does AI guilt affect students' willingness to use AI tools for different types of academic tasks?
4. What guidelines and educational policies should be developed to address AI guilt and support responsible AI use in schools?

**Methodology**
The study on AI guilt among secondary school students was developed through a structured multi-step process. It began with an extensive literature review covering existing research on AI usage, including the Technology Acceptance Model (TAM) by Davis (1989) and the Unified Theory of Acceptance and Use of Technology (UTAUT) developed by Venkatesh et al. (2003). Additionally, concepts related to academic integrity, Impostor Syndrome, and cognitive dissonance were explored.

To ensure the questions were comprehensive and relevant, the author consulted with several experts. Based on the literature review and expert discussions, open-ended questions were designed to capture the depth and complexity of students' experiences and perceptions regarding the use of AI tools in academic settings, particularly focusing on AI guilt.

The goal was to explore the viewpoints and personal experiences that might lead to AI guilt among secondary school students. These questions were designed to delve into the participants' concerns, personal experiences, and suggestions for mitigating AI guilt. By using open-ended questions, the study aimed to uncover rich, detailed narratives that provide deeper insights into the social and psychological impacts of AI tool usage among students.

Three open-ended questions were formulated to guide the qualitative inquiry as shown in table 1.

| | |
|---|---|
| Open1 | What are your main concerns about using AI tools in your work? |
| Open2 | Can you describe a situation where you felt guilty using an AI tool for your work? |
| Open3 | Do you believe ethical guidelines could help mitigate AI guilt? If so, what should they include? |

Table 1: Open-ended questions of Student's AI Guilt

**Sample Size and Demographics**
The AI Guilt questionnaire with the three open-ended questions, available in both English and Chinese versions, was administered to secondary school students at a Teaching and Learning AI EdTech Exhibition in Hong Kong. This exhibition provided an ideal setting for gathering data from a diverse group of students actively engaged with AI tools in their educational contexts. A total of 97 students participated in the study.

The demographic breakdown of the participants included 37 female students and 57 male students, with 3 students not specifying their gender. The intended age range varied from 15 to 18 years, with an average age of 16.5 years. The participants came from various educational backgrounds, offering a comprehensive perspective on the use of AI tools across different stages of secondary education.

**Results**
**Data Analysis of AI Guilt**
The data collected from the AI Guilt questionnaire were analyzed using qualitative content analysis. Transcriptions of the students' open-ended responses were coded inductively to identify significant concepts and ideas. These initial codes were then grouped into broader themes through thematic analysis, revealing key dimensions of factors that lead to AI guilt such as (1). Perceived laziness and authenticity (2). Fear of judgment, and (3). Concerns about identity and self-efficacy. The coding process was independently reviewed by multiple researchers to ensure reliability and validity, and discrepancies were resolved through consensus. The final themes were interpreted in the context of theoretical frameworks, to link students' experiences of AI guilt to broader constructs of technology acceptance, social influences, and educational practices. This comprehensive approach provided deep insights into the social and psychological impacts of AI tool usage among secondary school students.

*Research Question 1: What are the prevalent factors that lead to AI guilt among secondary school students?*

For this question, the qualitative data analysis followed an inductive approach, aiming to uncover emergent themes from the participants' responses. This method allowed for a grounded understanding of the students' perceptions of AI guilt. This process generated a wide array of codes representing various aspects of students' experiences. During the coding process, it became evident that certain themes consistently emerged across multiple responses. These included concerns about perceived laziness, fear of judgment, and issues related to identity and self-efficacy. To provide a structured and coherent analysis, the overlapping themes were organized into these three main dimensions and within each of these main dimensions, specific sub-categories were identified to capture varied aspects of students' experiences. For example, under perceived laziness or inauthenticity, sub-categories included concerns about taking shortcuts and not fully engaging with learning material.

**Perceived Laziness or Inauthenticity**
The perception of laziness or inauthenticity is a significant factor contributing to students' feelings of guilt when using AI tools. This guilt arises from the belief that utilizing AI for academic work shortcuts the learning process, thereby undermining the authenticity of their efforts and outputs. This sentiment is closely related to the concept of impostor syndrome, where individuals feel fraudulent or unworthy of their achievements.

*Ethical Dilemmas and Perceived Cheating*
Students often equate the use of AI with academic dishonesty. The data shows numerous instances where students explicitly express feelings of guilt associated with perceived cheating. For example, one student stated, "I feel very guilty. I copied and pasted a question and put it into ChatGPT to help me do it." This direct admission of guilt reflects a deep-seated concern about ethical conduct. Similarly, another student mentioned, "For my English writings I have used AI to write some paragraphs for me. It made me feel like I was cheating and wasn't learning anything." This reflects a broader concern that using AI tools equates to taking an unfair advantage, which not only breaches academic integrity but also diminishes the personal educational experience.

*Compromised Learning and Skill Development*
The perception that AI use might lead to laziness is also rooted in concerns about skill development. Students worry that over-reliance on AI will prevent them from fully engaging with their studies and developing essential academic skills. One student articulated this concern by saying, "That by using AI I will not be able to learn how to do these things on my own and my skills will not develop." This highlights a fear that AI tools may enable students to bypass critical learning processes, leading to superficial understanding rather than deep, conceptual learning.

Another student remarked, "When I use it for last-minute work and it's not original," indicating that using AI to meet deadlines without personal effort feels like cutting corners. This sentiment suggests that students recognize the potential for AI to facilitate procrastination and reduce the quality of their learning experiences. The reliance on AI for quick fixes rather than engaging with the material more deeply can foster a sense of laziness, further entrenching feelings of guilt.

*Authenticity and Originality in Academic Work*
Students also express concerns about the authenticity of their academic work when using AI tools. Many feel that AI-generated content lacks the personal touch and originality that characterize genuine student effort. For instance, one student mentioned, "Making my work less authentic. When I do my homework with AI." This statement underscores the belief that AI use can result in work that does not fully reflect the student's own knowledge, ideas, or effort.

Similarly, another student stated, "I worry that it'll mean my work isn't original. Also, since I'm considering an art route for my career as well, I fear that AI would replace me." This response highlights a specific anxiety in creative fields, where originality, creativity and personal expression are paramount. The fear that AI could produce work indistinguishable from human-created content can make students feel that their contributions are less valuable or meaningful.

*Impostor Syndrome and Psychological Impact*
The feelings of fraudulence associated with using AI tools are reminiscent of impostor syndrome, where individuals doubt their accomplishments and fear being exposed as "frauds." This psychological phenomenon is evident in student responses that reflect a sense of unworthiness or dishonesty. For example, a student noted, "I feel like I am not being truthful when I use it," which mirrors the internal conflict seen in impostor syndrome. This internal conflict can lead to significant stress and anxiety, as students grapple with the discrepancy between their actions (using AI) and their self-perception as honest, hardworking individuals.

Perceived laziness or inauthenticity is a profound factor driving students' guilt when using AI tools. This guilt is rooted in ethical dilemmas, concerns about compromised learning and skill development, fears about the authenticity and originality of their work, and the psychological impact akin to impostor syndrome.

**Fear of Judgment**
Fear of judgment is another significant factor contributing to the guilt students feel when using AI tools. This fear arises from concerns about how their use of AI will be perceived by peers, educators, and society at large. Students worry that reliance on AI might be seen as a shortcut or a sign of incompetence, which can lead to feelings of inadequacy and guilt.

*Anxiety Over Educators' Perceptions*
Many students express anxiety about how their teachers will perceive their use of AI tools. This concern is rooted in the fear of being judged as lazy, dishonest, or lacking in intellectual effort. For instance, one student stated, "I am scared that the teachers will scold me when I use AI," highlighting the fear of punitive reactions from educators. Another student echoed this sentiment by saying, "Teacher might think I'm cheating," which underscores the worry that teachers will view AI use as a form of academic dishonesty rather than a legitimate tool for learning and research.

This fear is not unfounded, as academic institutions often emphasize the importance of originality, critical thinking, and personal effort. The use of AI can conflict with these academic values, leading students to feel that their educators will not only disapprove of their use of AI but may also question their academic integrity and commitment to learning.

*Peer Judgment and Social Stigma*
Beyond educators, students are also concerned about how their peers will perceive their use of AI tools. The fear of judgment from peers can be equally, if not more, distressing. Students worry that admitting to the use of AI might lead to social stigma or diminished respect among their classmates. This is particularly significant in a collaborative academic environment where peer perceptions can influence one's academic reputation and social standing.

For example, one student mentioned the concern of being viewed negatively by peers: "When others (peers) give me work and after I use AI to do it, they said I did good," indicating a complex dynamic where the approval or disapproval of peers plays a critical role in their feelings of guilt.

*Reduced Truthfulness and Authenticity*
The fear of judgment can also lead students to be less truthful about the extent to which they use AI in their work. This reduction in transparency is driven by the desire to avoid negative judgment and maintain a facade of academic integrity. Students might downplay or hide their use of AI tools to conform to the expectations of originality and effort prevalent in academic settings.

Would students frame their AI usage in ways that are more acceptable to themselves and others, thereby reducing the perceived need for truthfulness? A student stated, "You will not feel guilty if you haven't been using AI for malicious purposes," of course, what is malicious purposes is still undefined. Students may have a tendency to justify AI use based on perceived intent rather than actual practice.

The fear of judgment is a multifaceted factor contributing to students' guilt when using AI tools. This fear encompasses concerns about how their use of AI will be perceived by educators, peers, and society, leading to anxiety about being judged as lazy, dishonest, or intellectually deficient. The fear of judgment also drives students to be less truthful about their AI usage, reducing transparency and authenticity in their academic work.

**Identity and Self-Efficacy Concerns**
Identity and self-efficacy concerns significantly contribute to the guilt students feel when using AI tools in their academic work. These concerns revolve around questions of personal capability, the authenticity of one's efforts, and the role of human agency in an increasingly automated world. As AI tools become more sophisticated and integrated into academic practices, students grapple with the implications for their self-worth, competence, and the value of their intellectual contributions.

*Impact on Personal Capability and Authenticity*
Students worry that relying on AI tools compromises the authenticity of their work and undermines their personal capability. This concern is evident in the way students perceive the role of AI in their learning processes. For instance, one student expressed, "That by using AI I will not be able to learn how to do these things on my own and my skills will not develop." This statement reflects a fear that AI use might prevent students from fully engaging with and mastering the material, leading to superficial understanding rather than deep learning.

Another student shared, "I feel like I am not being truthful when I use it," indicating a belief that AI-generated work lacks the personal touch and genuine effort required for true learning. This perception that AI usage leads to inauthentic work diminishes the student's sense of accomplishment and authenticity, making them feel that their academic outputs are not genuinely theirs.

*Human Agency vs. Machine Efficiency*
The balance between human agency and machine efficiency is another critical aspect of identity and self-efficacy concerns. Students are often caught between the efficiency and capabilities of AI tools and their desire to maintain control over their learning processes. The ease with which AI can perform complex tasks, such as writing essays or solving problems, can lead students to question the value of their own efforts. The perception that AI can perform tasks more effectively and efficiently than humans can lead to a devaluation of personal effort. Students might feel that their skills and knowledge are less valuable when compared to the capabilities of AI tools, one student mentioned "I will never be better than AI".

Another student articulated this dilemma by saying, "Developing over-reliance. When I use it for last-minute work and it's not original." This highlights the tension between utilizing AI for efficiency and maintaining a sense of personal agency and originality in their work. The student's admission of using AI for last-minute tasks suggests a conflict between the convenience of AI and the desire to produce work that truly reflects their abilities and effort.

*Cognitive Dissonance and Academic Integrity*
Cognitive dissonance theory, as proposed by Festinger (1957), provides a useful framework for understanding the discomfort students feel when their actions conflict with their beliefs about originality and personal effort. When students use AI tools, they may experience cognitive dissonance if they believe that true academic integrity requires human-generated effort and creativity. This dissonance arises because the use of AI contradicts their internalized values about learning and achievement.

For example, a student might feel proud of an essay generated with significant AI assistance but simultaneously guilty because it does not represent their independent effort. This internal conflict can lead to significant stress and anxiety, as students struggle to reconcile their actions with their beliefs about what constitutes genuine academic work. As one student noted, "When I write an essay for an unimportant yet time-consuming task (not graded)," they felt that using AI was acceptable, but this sentiment changes when the stakes are higher, indicating an underlying belief in the importance of personal effort and originality.

*Fear of Future Implications*
Students also worry about the long-term implications of AI on their identity and self-efficacy. As AI tools become more integrated into various aspects of life and work, students fear that their reliance on these tools might diminish their ability to succeed independently. One student noted, "AI tools will take over jobs, and less humans would work," reflecting a broader anxiety about the potential for AI to render human skills obsolete. This fear can lead to a diminished sense of self-efficacy, as students question the value of their efforts in a future dominated by AI.

| Students' Prevalent Factors to AI Guilt | | |
|---|---|---|
| Perceived Laziness or Inauthenticity | Fear of Judgment | Identity and Self-Efficacy Concerns |
| *Ethical Dilemmas and Perceived Cheating* | *Anxiety Over Educators' Perceptions* | *Impact on Personal Capability and Authenticity* |
| *Compromised Learning and Skill Development* | *Peer Judgment and Social Stigma* | *Human Agency vs. Machine Efficiency* |
| *Authenticity and Originality in Academic Work* | *Reduced Truthfulness and Authenticity* | *Cognitive Dissonance and Academic Integrity* |
| *Impostor Syndrome and Psychological Impact* | | *Fear of Future Implications* |

Table 2: Students' Prevalent Factors to AI Guilt

Identity and self-efficacy concerns are critical factors contributing to students' guilt when using AI tools. These concerns involve the impact on personal capability and authenticity, the balance between human agency and machine efficiency, cognitive dissonance related to academic integrity, the devaluation of personal effort, and fears about future implications. Table 2 shows the prevalent factors to students' AI guilt.

> ***AI Guilt** as a psychological phenomenon wherein individuals feel guilt or moral discomfort when use AI tools and feel that their reliance on these technologies might be perceived negatively by others or feel disingenuous. This could be particularly relevant in academic, professional, or creative contexts where originality and individual effort are highly valued.*

Based on the findings, AI guilt may arise from:

- Perceived Laziness or Inauthenticity: Concerns that using AI is a form of cheating or a shortcut that diminishes the value or authenticity of the student's output. The use of AI in academic work can sometimes be perceived as a form of cheating, where students feel they are gaining an unfair advantage or not fully engaging with the learning material. This perception can lead to significant guilt and ethical concerns. This is similar to Clance and Imes' Impostor syndrome (1978). The syndrome and its relevant construct measures students' feelings of frauds when they attribute their academic success to AI assistance rather than their own abilities.
- Fear of Judgment: Anxiety over how others (peers, supervisors, society at large) will perceive their use of AI, particularly in terms of ethics, creativity, or intellectual effort. Academic institutions often emphasize originality, critical thinking, and personal effort. The use of AI can conflict with these norms, leading to feelings of guilt among students who use AI tools in their academic work.
- Identity and Self-Efficacy Concerns: Questions about personal capability and the role of human agency when machines can perform complex tasks effectively and efficiently. In educational settings, students may feel that using AI tools to assist with writing assignments, research, or problem-solving compromises the authenticity of their work. This guilt arises from a belief that true learning and academic integrity require original, human-generated effort. Studies on cognitive dissonance, such as Festinger's (1957) theory, provide a useful lens for examining the discomfort students feel when their actions (using AI tools) conflict with their beliefs about originality and personal effort. As AI tools become more sophisticated, there is a fear that students' own skills and knowledge may be devalued. The ease with which AI can perform certain tasks might make students question the worth of their own efforts and learning processes.

*Research Question 2: What are the social and psychological impacts of AI guilt on secondary school students?*

**Social and Psychological Impacts of AI Guilt on Secondary School Students**
The data reveals that AI guilt exerts significant social and psychological impacts on secondary school students. These impacts manifest in various ways, influencing students' mental health, peer relationships, and overall academic experience.

*Psychological Stress and Anxiety*

One of the most profound psychological impacts of AI guilt is the heightened stress and anxiety students experience. The internal conflict between the convenience of using AI and the desire to maintain academic integrity can be mentally taxing. For example, a student expressed, "I used AI for my homework and got almost full mark and I felt guilty knowing it isn't my work." This guilt is compounded by the constant fear of being caught or judged, leading to increased anxiety levels. The mental burden of worrying about whether their use of AI is ethical or will be perceived as cheating can create a persistent state of unease, affecting their overall well-being and academic performance.

*Loss of Self-Efficacy*
AI guilt also undermines students' self-efficacy, or their belief in their own ability to succeed. When students rely on AI tools for tasks they feel they should be able to complete independently, it can lead to a diminished sense of their own capabilities. One student noted, "That I may lose my creativity and completely depend on AI." This sentiment reflects a broader fear that dependence on AI will prevent them from developing essential skills and knowledge, leading to a lower sense of self-worth and academic confidence. The belief that they are not genuinely learning or achieving on their own can have long-term effects on their educational motivation and self-esteem.

*Peer Relationships and Social Stigma*
The social impact of AI guilt is evident in the way it affects peer relationships and the potential for social stigma. Students are often concerned about how their use of AI tools will be perceived by their classmates. For instance, a student mentioned, "the teacher might find out," indirectly suggesting a similar concern about peers thinking the same. This fear can lead to social isolation, as students may avoid discussing their use of AI tools with friends or classmates to prevent negative judgment. The desire to fit in and be accepted by peers can make students more reluctant to use AI, even when it could be beneficial, further exacerbating feelings of guilt and anxiety.

*Fear of Academic Repercussions*
Students also experience a pervasive fear of academic repercussions as a result of AI guilt. The worry that their teachers might discover their use of AI and penalize them can lead to a significant psychological burden. For example, one student stated, "I worry teacher deduct my mark." This fear can deter students from fully utilizing AI tools, even in cases where they might enhance learning and understanding. The constant anxiety about potential academic consequences can distract students from their studies and contribute to a negative overall school experience.

The social and psychological impacts of AI guilt on secondary school students are multifaceted and significant. These impacts include heightened stress and anxiety, erosion of self-efficacy, strained peer relationships, fear of academic repercussions, ethical dilemmas, and long-term implications for learning and development.

*Research Question 3: How does AI guilt affect students' willingness to use AI tools for different types of academic tasks?*

**Impact of AI Guilt on Students' Willingness to Use AI Tools for Academic Tasks**
The phenomenon of AI guilt significantly influences students' willingness to utilize AI tools across various academic tasks. The data reveals a complex relationship where the type of task, the perceived ethical implications, and the potential impact on learning and skill development all play crucial roles in determining whether students feel comfortable incorporating AI into their academic work.

*Research and Idea Generation*
Students appear more willing to use AI tools for tasks that are perceived as less directly tied to their homework, such as research and idea generation. This is because these tasks often involve gathering information and brainstorming, activities where AI is seen as an aid rather than a substitute for personal effort in students' perspectives. For example, one student mentioned, "I use it to give me some ideas," indicating a comfort level with using AI to enhance their creativity and research capabilities. The willingness to use AI for such tasks is also linked to the perception that these tools can augment the learning process without compromising the authenticity of the final output. This suggests that when AI tools are framed as supportive aids that enhance rather than replace human effort, students are more likely to use them without feeling guilty.

*Writing and Homework Assignments*
Conversely, students exhibit significant hesitation in using AI tools for writing and completing homework assignments, where the ethical stakes are perceived to be higher. The fear of being accused of cheating and the internal conflict over authenticity contribute to this reluctance. One student shared, "For my English writings I have used AI to write some paragraphs for me. It made me feel like I was cheating and wasn't learning anything." This response highlights the guilt associated with using AI to produce substantive content, as students feel that this undermines their personal engagement with the material and their development of writing skills. The data indicates that while students might use AI tools to aid in the initial stages of writing, such as drafting outlines or generating ideas, they are cautious about relying on AI for the actual composition to avoid feelings of inauthenticity and guilt.

*Problem-Solving and Technical Tasks*
When it comes to problem-solving and technical tasks, such as those in mathematics and science, students' willingness to use AI tools varies significantly. Some students feel that AI can be a valuable tool for understanding complex concepts and checking their work, which can enhance their learning experience. For example, a student noted, "I don't use AI to complete assignments fully, only to teach myself concepts better." This indicates a strategic use of AI to supplement their understanding without relying on it entirely, thereby maintaining a balance between leveraging technology and ensuring personal effort. However, there is still a pervasive fear that over-reliance on AI for problem-solving can erode essential analytical and critical thinking skills, leading to a cautious approach in its application.

AI guilt profoundly affects students' willingness to use AI tools for different types of academic tasks. While there is a greater acceptance of AI in research and idea generation, where the tools are seen as supportive aids, there is significant hesitation in using AI for writing, homework assignments, and creative projects due to concerns about academic integrity, authenticity, and personal skill development. For problem-solving and technical tasks, students adopt a more balanced approach, using AI to enhance understanding without fully relying on it. In the discussion, we will explain the issues of such thinking.

***Research Question 4: What guidelines and educational policies should be developed to address AI guilt and support responsible AI use in schools?***

**Ethical Guidelines to Mitigate AI Guilt**
The responses to open-ended question 3, which asked whether students believe ethical guidelines could help mitigate AI guilt and what those guidelines should include, reveal a nuanced understanding of the potential benefits and limitations of such guidelines. From the students' responses, it appears that many schools in Hong Kong do not have specific guidelines regarding the use of AI tools. Consequently, many students are uncertain about whether they are permitted to use AI, and if so, to what extent it is appropriate for their schoolwork. This uncertainty is evident in the section on Research and Idea Generation. From an academic integrity perspective, research and idea generation should not be replaced by AI. However, students' perspectives on AI ethics are varied and uncertain.
Our findings also highlight several key themes, indicating a general consensus that well-crafted guidelines could significantly reduce feelings of guilt associated with using AI tools.

*Need for Clear and Specific Guidelines*
Many students express a belief that clear and specific ethical guidelines are essential to mitigate AI guilt. These guidelines should delineate the appropriate and inappropriate uses of AI tools in academic work. For instance, one student suggested, "Yes. Give rules on what parts of the process to use AI, e.g., in the research phase." This indicates a need for detailed instructions on how AI can be integrated into different stages of academic work without compromising the integrity of the learning process. By specifying when and how AI tools can be used, such guidelines can help students navigate their use ethically, ensuring they do not overstep boundaries that might lead to academic dishonesty or diminished learning. This would also reduce student anxiety.

*Encouraging AI as a Supplementary Tool*
Several students highlighted the importance of using AI as a supplementary tool rather than a replacement for personal effort. For example, one student mentioned, "Don't use AI in a way that would compromise instead of enhance your learning experience." This suggests that guidelines should emphasize the role of AI in supporting human capabilities, such as providing inspiration, aiding in research, or offering new perspectives, rather than doing the work for the student. By framing AI as a tool that supports and enhances learning rather than substitutes for it, guidelines can help students use AI in ways that preserve the authenticity and originality of their work.

*Limiting Direct AI Output Usage*
Another recurring theme is the restriction of direct usage of AI-generated content in final submissions. Students suggest that guidelines should prevent the direct copying of AI outputs. For instance, one student proposed, "Yes. They (guidelines) should include that AI should only be used to get ideas but not to copy." This approach would ensure that while students can benefit from AI's capabilities in generating ideas or providing initial drafts, the final output should always be a product of their effort and understanding. Such a rule reinforces the importance of personal engagement and intellectual effort, thereby reducing guilt associated with perceived laziness or inauthenticity.

*Ethical Use and Honesty*
Honesty and transparency about AI usage are also crucial components of ethical guidelines, according to student responses. One student emphasized, "Yes, they should include guidelines on how to use it responsibly." Responsible usage includes being honest about the extent of AI assistance and ensuring that AI is used ethically. This might involve properly citing AI-generated content, clearly indicating which parts of the work were assisted by AI, and ensuring that AI is used in a way that aligns with academic integrity principles. By promoting honesty and ethical behavior, guidelines can help students feel more confident and less guilty about using AI tools.

*Educational Integration and Support*
Several students also pointed out the need for educational institutions to integrate these guidelines into their teaching practices and provide adequate support for students. One student noted, "Yes, school giving a talk," suggesting that educational workshops or discussions about the ethical use of AI could be beneficial. Such initiatives could help demystify AI tools, educate students on ethical usage, and provide a platform for addressing concerns and questions. This proactive approach by educational institutions can help normalize the use of AI as a legitimate tool in academia, reducing the stigma and guilt associated with its use.

The analysis of responses indicates that students generally believe ethical guidelines could help mitigate AI guilt, provided these guidelines are clear, specific, and well-integrated into the educational framework. Key components of these guidelines should include: specifying appropriate uses of AI, encouraging AI as a supplementary tool, limiting the direct use of AI outputs, promoting honesty and transparency, and integrating educational support to foster a better understanding of AI tools. By addressing these areas, ethical guidelines can help students use AI tools responsibly and confidently, thereby reducing feelings of guilt and enhancing the overall educational experience.

**Discussion: Student Perception of AI Guilt**

The rise of AI in education is both a marvel and a conundrum. As AI tools become more sophisticated, their integration into academic settings is inevitable. However, this integration has not been seamless, giving rise to a phenomenon termed "AI guilt." This guilt arises from the moral discomfort users feel when relying on AI tools for tasks traditionally performed by humans. While the ethical implications of AI use in education are still being debated, it is clear that AI guilt reflects deeper concerns about originality, authenticity, and the intrinsic value of human effort.

At its core, AI guilt is about more than just cheating or taking shortcuts; it's about the existential crisis that arises when machines encroach upon areas traditionally dominated by human intellect and creativity. The findings with students expressing concerns about AI use equating to cheating or taking shortcuts. The fear of being perceived as lazy or inauthentic is rooted in a broader societal narrative that equates hard work with moral virtue. In this context, using AI feels like undermining one's moral and intellectual worth. Historically, education has been a journey of personal and intellectual growth, characterized by overcoming challenges through individual effort. The integration of AI challenges this paradigm by introducing a tool that can perform complex tasks with minimal human intervention. **This shift forces us to reconsider what we value in education: Is it the process of learning, or the final output?** If AI can enhance learning outcomes, should we still prioritize traditional methods that emphasize effort over efficiency?

The feelings of guilt associated with AI use can be compared to impostor syndrome (Clance & Imes, 1978), where students feel fraudulent because their achievements are assisted by AI rather than being the result of their own efforts. This psychological phenomenon reflects a deeper cognitive dissonance where students' actions (using AI) conflict with their beliefs about originality and personal effort (Festinger, 1957). The moral distress can lead to a cycle of guilt and self-doubt, where students continually question the righteousness of their actions and their adherence to ethical standards. AI tools, while enhancing efficiency, also challenge students' perceptions of their capabilities. The ease with which AI can perform certain tasks may lead students to doubt their own skills and worth, the dilemma between machine's efficiency and self-efficacy. This can result in a diminished sense of self-efficacy, as students may feel that their intellectual contributions are devalued in comparison to what AI can achieve.

The **traditional concept of academic integrity**, which emphasizes originality and personal effort, may **need to be redefined in the age of AI**. If AI tools can enhance learning and provide significant educational benefits, **it may be necessary to develop new ethical frameworks** that balance the use of AI with the **preservation of academic values**.

Educators play a crucial role in shaping how AI tools are perceived and used in academic settings. By fostering an environment that values ethical AI use and personal effort, educators can help mitigate AI guilt. This involves creating clear guidelines that delineate acceptable AI use, encouraging transparency, and promoting discussions about the ethical implications of AI in education. **The findings clearly indicate that students are confused about what AI should and should not be used for.** The fact that many students believe generating ideas using AI should be considered genuine use is alarming. **If all ideas are generated by AI, it will lead to a homogenization of thoughts** (Chan & Colloton, 2024, p. 56), resulting in limited perspectives. Redesigning assessments to evaluate students' processes and skills will be crucial in the future. As AI continues to evolve, it is essential to prepare students for a future where AI will be an integral part of various professions. This preparation involves not only teaching students how to use AI tools effectively but also instilling a strong ethical foundation that guides their use of technology. By doing so, we can ensure that students are equipped to navigate the complexities of AI use in their professional and personal lives.

Developing comprehensive ethical guidelines for AI use in education is crucial. These guidelines should specify the appropriate uses of AI, emphasize the role of AI as a supplementary tool, and promote honesty and transparency. By providing clear and specific instructions, these guidelines can help students navigate the ethical landscape of AI use, reducing feelings of guilt and uncertainty. Encouraging open discussions about AI and its implications can help demystify the technology and reduce the stigma associated with its use. These discussions can provide a platform for students to express their concerns, share their experiences, and develop a more nuanced understanding of AI's role in education.

To mitigate AI guilt, it is important to balance the use of AI tools with personal effort. Educators can encourage students to use AI for supportive tasks while ensuring that the final output reflects the student's own efforts and understanding. This approach can help maintain the authenticity and integrity of students' work while leveraging the benefits of AI. Long-term psychological impacts of AI guilt extend beyond immediate academic concerns, potentially affecting students' overall development and approach to learning. The fear of becoming overly reliant on AI can lead students to avoid using valuable technological tools altogether, potentially hindering their ability to adapt to future technological advancements. This avoidance behavior, driven by guilt and fear, can ultimately restrict their opportunities for growth and learning in an increasingly digital world.

The broader historical context of technological adaptation offers additional insights into AI guilt. Throughout history, major technological advancements have elicited fears of obsolescence and ethical concerns, from the Industrial Revolution to the rise of the internet (Brynjolfsson & McAfee, 2014). These transitions often involve initial resistance followed by eventual integration, suggesting that current discomfort with AI might similarly evolve as mentioned by many student responses. However, the unprecedented pace of AI advancement necessitates a more rapid and comprehensive adaptation strategy (Raupp, 2017; Oxford University Press, 2023).

**Conclusion**

AI guilt represents a significant challenge in the integration of AI in education, reflecting deeper concerns about originality, authenticity, and the value of human effort. By addressing these concerns through ethical guidelines, open discussions, and a balanced approach to AI use, we can create an educational environment that embraces the benefits of AI while preserving the core values of learning. As we navigate this new ethical frontier, it is essential to continuously reflect on and adapt our educational practices to ensure that they align with both technological advancements and our intrinsic values. In doing so, we can prepare students for a future where AI is not just a tool, but a partner in their intellectual and personal growth.

*Future Directions*
The study on AI guilt among secondary school students provides a foundational understanding of the emotional and ethical implications of AI usage in academic settings. Future research should expand on this work by exploring the long-term impacts of AI guilt on student development and academic integrity. Longitudinal studies could track students over several years to observe how their perceptions of AI evolve and how AI guilt influences their learning outcomes and career choices. Additionally, it would be beneficial to conduct cross-cultural studies to determine if AI guilt manifests differently in various educational systems and cultural contexts. This could provide a more global perspective on the phenomenon and help in developing universally applicable ethical guidelines.

Another important direction for future research is to investigate the effectiveness of interventions designed to mitigate AI guilt. Experimental studies could test different educational strategies, such as workshops on ethical AI use, integration of AI ethics into the curriculum, and the development of AI usage policies that emphasize transparency and personal effort. These studies could assess how these interventions impact students' perceptions of AI and their feelings of guilt. Moreover, research could explore the role of educators in shaping students' attitudes towards AI, examining how teachers' approaches to AI integration influence student experiences and ethical considerations.

*Limitations of the Study*
Despite its valuable insights, this study has several limitations that should be addressed in future research. First, the sample size of 97 students, though providing a preliminary understanding, is relatively small and may not be representative of the broader student population. Larger and more diverse samples are needed to generalize the findings more confidently. Additionally, the study was conducted in a specific educational context at a Teaching and Learning AI EdTech Exhibition in Hong Kong, which might influence the results due to the participants' heightened exposure to AI technologies. Future studies should include a wider range of educational settings to capture a more comprehensive picture of AI guilt.

Another limitation is the reliance on self-reported data, which may be subject to social desirability bias. Students might underreport their use of AI tools or their feelings of guilt to align with perceived social norms. Incorporating objective measures of AI usage and triangulating self-reported data with teacher assessments or peer evaluations could provide a more accurate depiction of AI guilt. Furthermore, the study primarily focuses on secondary school students; expanding the research to include primary school students, college students, and adult learners could offer insights into how AI guilt varies across different age groups and educational stages.

Lastly, while the study identifies key dimensions of AI guilt, such as perceived laziness, fear of judgment, and identity concerns as the prevalent factors, it does not delve deeply into the psychological mechanisms underlying these feelings. Future research should employ qualitative methods, such as in-depth interviews or focus groups, to explore the cognitive and emotional processes that contribute to AI guilt. Understanding these underlying mechanisms can inform more targeted interventions to support students in navigating the ethical landscape of AI in education.

**Declaration of generative AI and AI-assisted technologies in the writing process**

Statement: During the preparation of this work the author(s) used [ChatGPT 3.5/ Poe] in the writing process in order to improve the readability and language of the manuscript. After using this tool/service, the author(s) reviewed and edited the content as needed and take(s) full responsibility for the content of the published article.